\documentclass[aps,prl,twocolumn,groupedaddress,showpacs]{revtex4-1}

\usepackage{dcolumn}
\usepackage{graphicx,amssymb,amsmath}
\usepackage{bm}
\usepackage{natbib}
\usepackage{epstopdf}
\begin{document}

\title{Spherical spin-glass--Coulomb gas duality: solution beyond mean-field theory}

\author{Shimul Akhanjee $^{1,2}$}
\email[]{shimul@riken.jp}
\author{Joseph Rudnick $^2$}
\email[]{jrudnick@physics.ucla.edu}
%\thanks{}

\affiliation{$^1$Condensed Matter Theory Laboratory, RIKEN, Wako, Saitama, 351-0198, Japan\\
$^2$Department of Physics, UCLA, Box 951547, Los Angeles, CA 90095-1547}

%Collaboration name if desired  (requires use of superscriptaddress
%option in \documentclass). \noaffiliation is required  (may also be

\date{\today}

\begin{abstract}

We present an alternate solution of a Gaussian spin-glass model with infinite ranged interactions and a global spherical constraint at zero magnetic field. The replicated spin-glass Hamiltonian is mapped onto a Coulomb gas of logarithmically interacting particles confined by a logarithmic single particle potential. The precise free energy is obtained by analyzing the Painlev\'e $\tau^{IV} [n]$ function in the $n\to 0$ limit. The large $N$ thermodynamics exactly recovers that of Kosterlitz, Thouless and Jones  \cite{ktjonesPRL}. It is hoped that the approach here can be extended to apply to systems beyond the spherical model, particularly those in which destabilizing terms lead to replica symmetry breaking.  

\end{abstract}


\maketitle

Spin-glass systems are an important component of the longstanding effort to understand the statistical mechanics of disordered systems. In particular,  spin-glass models exemplify the interplay of frustration---an outgrowth of quenched randomness---and thermal fluctuations \cite{binderyoung}. Attention has centered on the phase transition from a high temperature paramagnetic phase to the spin-glass phase, in which the spins are frozen in random directions over macroscopic time scales \cite{edanderson,sherrkirkPRL,parisiPRL,chaksingh,bhattyoung}. 

Several decades ago, an alternative yet simple, non-Ising spin-glass model was proposed by Kosterlitz, Thouless and Jones  (KTJ) based on the exactly solvable spherical model (SM) \cite{ktjonesPRL}, which has been termed the spherical spin-glass model  (SSGM). KTJ constructed a solution based on the known methodology of the SM that depends on the density of eigenvalues of the exchange coupling matrix \cite{baxter}. For the SSGM, the exchange coupling matrix in the large $N$ limit is effectively a Wigner-Dyson random matrix that follows a semi-circle law distribution. One of the most appealing features of the KTJ analysis was its prediction of a spin-glass transition without the need to address replica symmetry breaking, as the quenched averaged free energy can be computed without requiring replicas at all. A second approach attempted to work within the replica framework, pointing out a Parisi like order parameter and possible replica symmetry breaking perturbations \cite{jag_rud,jag_eva_rud}.

In this letter, we introduce a third approach that solves the SSGM without recourse to the large $N$ random matrix analysis of KTJ. The SSGM is mapped onto an effective Hamiltonian of logarithmically interacting charges in a logarithmic single particle potential, the number of replicas corresponding to the number of particles $n$. This approach does not require any assumptions about the replica structure, and the $n\to 0$ limit can be extrapolated using inductive diagrammatic arguments to connect the grand canonical and canonical ensembles. The partition function is shown to be equivalent to the Painlev\'e $\tau^{IV}[n]$ function, which is solved explicitly to yield a  spin-glass transition that is in complete agreement with the results of KTJ. It is hoped that the spin-glass--Coulomb gas connection and the methods described here are general enough to be applied to solve both the Sherrington-Kirkpatrick and Edwards-Anderson models beyond the limitations imposed by existing mean-field and numerical approaches. We note that a general relationship between replica based Hamiltonians and Painlev\'e transcedents has been rigorously demonstrated \cite{kanzieperPRL2002} and therefore the approach can be possibly extended to the broader category of replica based or disorder-induced critical phenomena.

The SSGM is based on a strictly Gaussian model, in which the spins take on continuous values in a specified range. The Hamiltonian for $N$ spins is
\begin{equation}
\mathcal{H}_{sg} = - \sum\nolimits_{i>j}J_{ij}S_iS_j + \Lambda \sum\nolimits_i S_i^2
\label{eq:s1}
\end{equation}
where the quantity $\Lambda$ is the Lagrange multiplier that enforces the spherical constraint, $\sum_{i=1}^N \left< S_i^2 \right> = N$ and each exchange coupling $J_{ij}$ is subject to a Gaussian distribution, $P (J_{ij}) \propto e^{-NJ_{ij}^2/2\bar{J}^2}$, with $\bar J = J/T$.
Since we are interested in the quenched average of the free energy, we apply the replica technique and consider the annealed average of the $n^{\rm th}$ power of the partition function $Z_n$. In the first step we integrate over the $J_{ij}$'s in the usual manner to yield an effective four-spin interaction that couples different replicas, with $a,b$ replica indices that range from 0 to $n$. Next, these terms can be decoupled by introducing an auxiliary field $\stackrel{\leftrightarrow}{Q}$, in a Hubbard-Stratonovich transformation,
\begin{equation}
Z_n = \exp \left[  -Q^{a b} \sum\nolimits_iS_i^{a}S_i^{b} - \frac{N}{2\bar{J}^2}  (Q^{a b})^2 - \Lambda \sum\nolimits_{i, a} S_i^{a \, 2}\right]
\label{eq:s5}
\end{equation}
We can then integrate out the $S_i^{a}$'s:
\begin{equation}
Z_n = \exp \left[ - \frac{N}{2} \mathop{\rm Tr} \ln \left (\stackrel{\leftrightarrow}{I} \Lambda + \stackrel{\leftrightarrow}{Q} \right) - \frac{N}{2 \bar{J}^2} \mathop{\rm Tr}  (\stackrel{\leftrightarrow}{Q}^2)\right]
\label{eq:s6}
\end{equation}
In the final manipulation, we have to integrate over the matrix $\stackrel{\leftrightarrow}{Q}$, which is real and symmetric. We are led to the evaluation of a partition function over the eigenvalues of $\stackrel{\leftrightarrow}{Q}$, $\lambda_i$ having the form, $\mathcal{Z}_n = \int \exp \left[  \mathcal{H}_{cg} (\lambda)\right] d \lambda_1 \ldots d \lambda_n$, where the Hamiltonian $\mathcal{H}_{cg}$ corresponds to a one-dimensional gas of particles with logarithmic interactions, which we term a Coulomb gas (CG) \cite{mehta}.
\begin{equation}
\mathcal{H}_{cg} (\lambda) = \sum\nolimits_i {V (\lambda_i)}  + A\sum\nolimits_{i > j} {\ln \left| {\lambda _i  - \lambda _j } \right|} 
\label{eq:cgham}
\end{equation}
with the single particle potential
\begin{equation}
V (\lambda) = \frac{N}{2} \ln \left[ \Lambda + \lambda \right] + \frac{N}{2 \bar{J}^2} \lambda^2
\label{eq:s8}
\end{equation}
For the rest of this paper, we fix $A=2$ at the unitary ensemble, as this allows us to adapt results now in the literature \cite{mehta,FandW,okamoto1981}. We believe that this substitution preserves the essential features of the model.

As a prelude to the full calculation of the partition function, we outline the development of a diagrammatic approach to its evaluation. In the replica formalism, the limit $n \rightarrow 0$ must be taken with care.  Our basic strategy is to first work within the canonical ensemble and observe the general the trend of its $n$ dependence in order to transform to the grand canonical ensemble, creating an adjustable $n$ that can be extrapolated to zero. A useful method for treating the logarithmic interaction is a virial expansion of $\mathcal{Z}_n $,

\begin{equation}
\begin{array}{l}
 \mathcal{Z}_n  = \frac{1}{{n!}}\int {d\lambda _1 d\lambda _2  \cdots d\lambda _n e} ^{ - H[\lambda _i ]}  \\ 
  = \frac{1}{{n!}}\int {d\lambda _1 d\lambda _2  \cdots d\lambda _n e} ^{ - V[\lambda _i ]}  \\ 
  + \frac{1}{{n!}}\int {d\lambda _1 d\lambda _2  \cdots d\lambda _n e} ^{ - V[\lambda _i ]} \left ( { - A\sum\nolimits_{i > j} {\ln \left| {\lambda _i  - \lambda _j } \right|} } \right) \\ 
  +  \cdots  \\ 
 \end{array}
\end{equation}
\begin{figure}
\centerline{\includegraphics[width=1.5in]{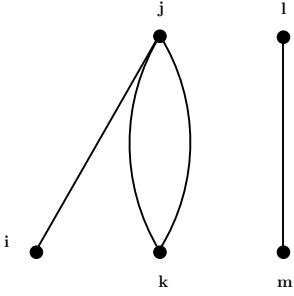}}
\caption{$L_5$, a particular 5-vertex diagram in the virial expansion of $\mathcal{Z}_n$}
\label{fig:5vertex}	
\end{figure}

Take for instance an arbitrary five vertex diagram, that we term $L_5$ as shown in Fig. \ref{fig:5vertex}. The expression to which this diagram corresponds to is given by:
\begin{equation}
\begin{array}{l}
 L_5  =  M_5 \int {d\lambda _i d\lambda _j d\lambda _k d\lambda _l d\lambda _m e^{^{ - V[\lambda _i ]} } } \left ( { - A\ln \left| {\lambda _j  - \lambda _i } \right|} \right) \times  \\ 
 e^{^{ - V[\lambda _j ]} } \left ( { - A\ln \left| {\lambda _k  - \lambda _j } \right|} \right)^2 e^{^{ - V[\lambda _k ]} }  \times  \\ 
 e^{^{ - V[\lambda _l ]} } \left ( { - A\ln \left| {\lambda _m  - \lambda _l } \right|} \right) e^{^{ - V[\lambda _m ]} }  \\ 
 \end{array}
\end{equation}
where the quantity $M_5  = n!  n_1 n_2 / ( (n - 5)!3!2!)$ is the overall weight of the diagram. Next, one accounts for the number of ways of assigning the $n$ total vertices to the 5 vertices in the diagram. Then, one multiplies by symmetry factors from permuting lines connected to a common vertex, yielding $n_1 = n_2 = 1/2$. One straightforwardly infers that any diagram is simply a product of expressions involving connected diagrams followed by the factor $\mathcal{Z}_{n-M}^{ (0)}$, where the ``bare'' partition function is given by,
$\mathcal{Z}_n^{ (0)}  = \frac{1}{{n!}}\left[ {\int {d\lambda e^{^{ - V[\lambda ]} } } } \right]^n $.

The grand partition function, $\Theta  (z) = \sum\nolimits_{n = 0}^\infty  {\mathcal{Z}_n z^n } $ can be formally constructed with each vertex multiplied by a fugacity factor $z$ such that $\Theta ^{ (0)}  (z) = \exp \left[ z q_0 \right]$, where $q_0  \equiv {\int {d\lambda e^{^{ - V[\lambda ]} } } } $. In the limit $n \rightarrow \infty$, $\Theta (z)$ and $Q_n$ satisfy the relations, $\ln \mathcal{Z}_n  = \ln \Theta  (z (n)) - nz (n)$ and $n = \frac{\partial }{{\partial z}}\Theta  (z)$. To extract the $n \rightarrow 0$ limit, the $n$ dependence of $L_5$ reduces to $n\times4! (-1)^4$. Dividing out $n$, and re-expressing the factorial in terms of a $\Gamma$ function, we are left with the remainder $-  ( - 1)^5 \int_0^\infty  {t^4 e^{ - t} dt}$.

Next we make use of the zeroth order, non-interacting term in the partition function, $q_0 ^n  \approx 1 + n\ln q_0 $. Since, $L_5$ scales as $1/q_{0}^5$, we can absorb this factor by recasting $L_5$ as, $-  ( - 1)^5 \int_0^\infty  {dt  (t^5 e^{ - q_0 t}/t)}$.
Finally, this allows the virial expansion to be completely factored in terms of its $n$ dependence. The integral form of the logarithm, $\ln q_0  = \int_0^\infty  {dt (e^{ - t}  - e^{ - q_0 t})/t }$ can be substituted with $e^{-q_0 t}=\Theta^{ (0)} (t)$, or the non-interacting grand partition function at negative fugacity. The disorder-averaged free energy is, then,
\begin{equation}
\left\langle {\ln \mathcal{Z}} \right\rangle = \int_0^\infty  {dt (e^{ - t}  - \Theta  (-t))/t} 
\label{eq:intpart}
\end{equation}
Additionally, if the partition function of the $n$-component system takes the form,
$\mathcal{Z}_n = \int {{e^{n ({x_1} + {x_2} + \cdots {x_k})}}f ({x_1}, \ldots {x_k})\prod\nolimits_{i=1}^k {d{x_i}} }$
then one can easily show that Eq. (\ref{eq:intpart}) is equivalent to the standard replica calculation 
\begin{equation}
\left\langle {\ln \mathcal{Z}} \right\rangle=\mathop {\lim }\limits_{n \to 0} \left ( {{\mathcal{Z}_n - 1}} \right)/n = {\left. {d{\mathcal{Z}_n}/dn} \right|_{n = 0}}.
\label{eq:replica}
\end{equation}
\begin{figure}
\centerline{\includegraphics[height=1.5in]{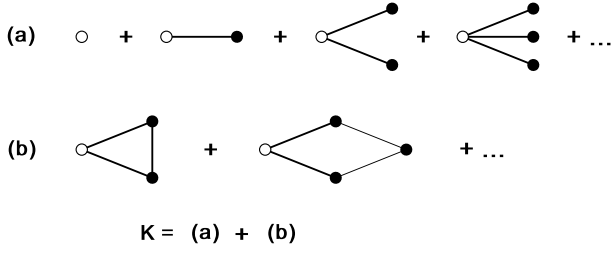}}
\caption{The full diagrammatic expansion of $K$. An empty circle stands for the vertex at point $x$ and the dark circles are at other points. A line is simply the distance function $\left| {x - x'} \right|$.  (a) The tree-level diagrams resulting in the mean-field density $\rho_{MF} (x)$ (b) Loop contributions beyond mean-field theory.}
\label{fig:tree}	
\end{figure}
Subsequently, the equation for the density can be represented as a diagrammatic series where $\rho  (x) = \rho_0 \times K$ and the graphical form of $K$ is shown in Fig. \ref{fig:tree}. $K$ can be generated by observing that $V (\lambda)$ couples to the density $\rho$ at the zeroth order, $\rho_0  (\lambda) = z e^{-V (\lambda)}$. The tree-level diagrams of Fig. \ref{fig:tree}  (a) result in a self-consistent series for mean field density $\rho_{MF} (x)$. The summation of these terms reduce to an inhomogeneous Sine-Gordon equation. The proper solution is non-trivial and a thorough analysis will be discussed in a separate article. In principle one should recover the known KTJ result:
\begin{equation}
{\mathcal{F}^{KTJ}} \simeq \left\{ \begin{array}{l}
  - \frac{{{J^2}}}{{4T}} - \frac{T}{4} \hspace{3cm}T<T_c\\ 
  - J + \frac{T}{2} + \frac{T}{2}\ln \left ( {\frac{J}{T}} \right)\hspace{1.4cm}T>T_c \\ 
 \end{array} \right.
 \label{eq:ktj}
 \end{equation}
 
We now leave our discussion of the mean field theory and focus on the full solution, which can be evaluated exactly using $\tau$ function theory of Painlev\'e systems.
The first task is to recast the Okamoto $\tau^{IV} [n]$ \cite{okamoto1981} integral discussed by Forrester and Witte  (F\&W) \cite{FandW} into a form that is equivalent to the partition function of Eq. (\ref{eq:cgham}), 
\begin{equation}
\begin{aligned}
&\tau^{IV}[n]= \frac{1}{C} \times \\
&\int_{-\infty}^t dx_1 \cdots \int_{-\infty}^t dx_n \prod_{j=1}^n e^{-x_j^2} (t-x_j)^{\mu} \prod_{1 \leq j <k \leq n}  (x_k-x_j)^2
\end{aligned}
\label{eq:FWint1}
\end{equation}

The integrand above can be rescaled by letting $x_j \rightarrow y_j \sqrt{Na}$, and let $t \rightarrow \Lambda \sqrt{Na }$, yielding 
\begin{equation}
\begin{aligned}
&\mathcal{Z}_n= (Na)^{ (n (n-1)/2+n+\mu)/2}\times \\
&\prod_{j=1}^{n}e^{-Nay_j^2}  (\Lambda-y_j)^{\mu} \prod_{1 \leq j<k \leq n} (y_k-y_j)^2
\label{eq:integ1}
\end{aligned}
\end{equation}
and the $n=1$ term is given by,
\begin{equation}
\tau^{IV}[1]=\int_{-\infty}^{\infty} (t-x)^{\mu} e^{-x^2} dx
\label{eq:n1int1}
\end{equation}
Recurrence relations for $\tau^{IV}[n]$ have been thoroughly investigated by F\&W  in the context of certain random matrix averages. These relations are well suited for our purposes,
\begin{equation}
\sigma^{IV}[n] = \mathop{\rm det} \left[\frac{d^{i+j}}{dt^{i+j}}\sigma^{IV}[1] \right]
\label{eq:sig1}
\end{equation}
where the determinant is of an $n \times n$ matrix with indices $(i,j)$ and the relationship between $\sigma^{IV}[n]$ and $\tau^{IV}[n]$ is given by,
\begin{equation}
\sigma^{IV}[n] = 2^{n (n-1)} \pi^{n/2}\left ( \prod_{j=1}^{n-1} j! \right) e^{x^2 n} \tau^{IV}[n]
\label{eq:sig5}
\end{equation}
It follows that the $n=1$ part 
\begin{equation}
\sigma^{IV}[1] = e^{x^2} \tau^{IV}[1]
\label{eq:sig2}
\end{equation} 
is used to generate higher orders in $n$ via Eq.(\ref{eq:sig1}). This allows us to extrapolate the $n$ dependence of $\tau^{IV}[n]$ in the case in which $\mu=-N/2$ for large postive $N$. Let's place $t$ slightly off the real axis and consider the integral of Eq. (\ref{eq:n1int1}) when $\mu=-1$, which can be shown to take the general form: 
\begin{equation}
\begin{aligned}
&\int_{-\infty}^{\infty} (x-t)^{-|\mu|} e^{-x^2} dx \\ 
&=  \frac{1}{ (|\mu|-1)!}\frac{d^{|\mu|-1}}{dt^{|\mu|-1}}  \left ( 2 \sqrt{\pi} e^{-t^2} \int_0^t e^{k^2} dk\right)
\end{aligned}
\label{eq:negmuint4}
\end{equation}
which is equivalent to the Dawson integral \cite{abramowitz}, 
\begin{equation}
e^{-t^2} \int_0^{t} e^{k^2} dk = \int_0^{\infty} e^{- k^2} \sin  (kt )\ dk
\label{eq:Dawson1}
\end{equation}
The evaluation of the derivatives of the integral as given by the R.H.S. of  (\ref{eq:Dawson1}) can be now taken, 
\begin{equation}
\begin{aligned}
&\lefteqn{\frac{d^{|\mu|-1}}{dt^{|\mu|-1}} \int_0^{\infty} e^{- k^2} \sin  (kt )\ dk} \nonumber \\ 
&= \frac{1}{2} \mathop{\rm Im} \left[ \int_0^{\infty}i^{|\mu|-1} \exp \left[ -k^2+ (|\mu|-1) \ln k + i k t \right] dk\right]
\label{eq:Dawson2}
\end{aligned}
\end{equation}
We can apply the stationary phase approximation and evaluate the integral by expanding about the extremum to yield the following result,
\begin{equation}
\begin{array}{l}
 {\tau ^{IV}}[1] = \sqrt {\frac{{\pi \left ( {t - \sqrt {{t^2} - 8 (|\mu|-1)} } \right)}}{{\sqrt {{t^2} - 8 (|\mu|-1)} }}}  \times  \\ 
 \exp \left[ {\frac{{t\sqrt {{t^2} - 8n} }}{8} +  (|\mu|-1)\ln \left ( {\frac{{\left ( {t - \sqrt {{t^2} - 8n} } \right)}}{4}} \right) - \frac{ (|\mu|-1)}{2} - \frac{{{t^2}}}{8}} \right] \\ 
 \end{array}
\label{eq:sp9}
\end{equation}
Equation (\ref{eq:sp9}) can be simplified by the functions $f (q)$ and $g (q)$, with $q\equiv t/\sqrt{|\mu|-1}\approx t/\sqrt{N}$ such that, 
\begin{equation}
\tau^{IV}[1]=g (q) \exp[Nf (q)]
\label{eq:sp10}
\end{equation}

We find empirically that the precise leading order in $n$ contribution to the determinant in  (\ref{eq:sig1}) yields,
\begin{equation}
\sigma^{IV}[n] =\left ( \prod_{j=1}^{n-1} j! \right) g (q)^n e^{Nnf_1 (q)} f_1^{\prime \prime} (q)^{n (n-1)/2}
\label{eq:sig3}
\end{equation}
where $f_1 (q) \equiv f (q)+q^2$. The final expression becomes,
\begin{equation}
\tau^{IV}[n] = e^{-t^2 n} 2^{-n (n-1)} \pi^{-n/2} g (q)^n e^{nNf (q)} f_1^{\prime \prime} (q)^{n (n-1)/2}
\label{eq:sig6}
\end{equation}
Taking the appropriate derivative of Eq. (\ref{eq:replica}) yields the $n\to0$ limit. The prefactor of Eq. (\ref{eq:integ1}) contains no interesting thermodynamic information; the key contributions to the free energy become
\begin{equation}
\begin{aligned}
&\mathcal{F}[\Lambda ] = \\
 &- T\left\langle {\ln \mathcal{Z}[\Lambda ]} \right\rangle  = T\left(Nf (y) + \ln \left[ {\frac{{2g (y)}}{{\sqrt {\pi f_1^{''} (y)} }}} \right] - {t^2}N \right)
 \end{aligned}
\end{equation}

The Lagrange multiplier can be eliminated by enforcing the spherical constraint via the relation $ - \frac{{\partial \left\langle {\ln \mathcal{Z}[\Lambda ]} \right\rangle }}{{\partial \Lambda }} = \sum\nolimits_{j = 1}^N {\left\langle {s_j^2} \right\rangle }  = N$, yielding:
\begin{equation}
 \frac{1}{4}  (N T /J) \left (-\sqrt{T^2 \Lambda^2 / J^2 -8 }+ T\Lambda /J \right)  =N
\label{eq:sphc1}
\end{equation}
having a solution of ${\Lambda_> } = 1 + \frac{{2{J^2}}}{{{T^2}}}$ where the range of $T$ lies above the branchpoint structure of the squareroot. However, for low enough values of $T$, there is no proper mathematical solution, and ${\Lambda _ < } = \frac{{\sqrt 8 J}}{T} + \mathcal{O} (1/N)$ in order to obey the spherical constraint. Hence, the complete free energy is given by,
\begin{equation}
\begin{array}{l}
\mathcal{ F} =  - T\left\langle {\ln \mathcal{Z}[{\Lambda _ > }]} \right\rangle  \\ 
  =  - \frac{NT}{8}\left ( { - \frac{{\left ( {2 + {{T'}^2}} \right)}}{{{{T'}^2}}}} \right. + T'\left ( {1 + \frac{2}{{T'}}} \right)\sqrt {{{T'}^2} + 4{{T'}^{ - 2}} - 4}  \\ 
 \left. { + 8\ln \left[ {\frac{{\sqrt N }}{{4T'}}\left ( {2 + T'\left ( {T' - \sqrt {{{T'}^2} + 4{{T'}^{ - 2}} - 4} } \right)} \right)} \right] - 4} \right) \\ 
 \end{array}
 \label{eq:fgreat}
\end{equation}
where $T'=T/J$. The critical temperature is obtained from the maximum of Eq. (\ref{eq:fgreat}), yielding $T_c=\sqrt 2 J$. After taking the large $N$ limit, we can compare the results to KTJ. Taking the appropriate derivatives of $\mathcal{ F}$ we have for the specific heat per site, 
\begin{equation}
{C_V}(T) = \left\{ \begin{array}{l}
 1\,\,\,\,\,\,\,\,\,\,\,\,\,\,\,\,\,\,\,\,\,\,\,\,\,\,\,\,\,\,\,\,\,\,\,\,T < {T_c} \\ 
 2 - {({T_c}/T)^2}\,\,\,\,\,\,\,\,T > {T_c} \\ 
 \end{array} \right.
\end{equation}
in precise agreement with KTJ, as the phase transition contains a discontinuous 3rd derivative or specific heat exponent $\alpha=-1$ similar to that observed in the three dimensional ideal Bose gas and the SM. Although our analysis is taken in the unitary ensemble  (A=2) rather than the orthogonal ensemble (A=1) appropriate to the model considered by KTJ, we expect no great discrepancies as we can generalize our results to account for the possibility of other ensembles by including an additional factor $B$ in the spherical constraint equation. Re-expressing Eq.(\ref{eq:sphc1}), 
\begin{equation}
BN = 2N{{T'}^2}\left[ {\Lambda  - \sqrt {{\Lambda ^2} - {{T'}^{ - 2}}} } \right]
\end{equation}
and solving it explicitly, we find that it is satisfied if $T' \ge B/2$ and therefore we expect that $B$ and subsequently the ensemble characteristics should only affect the value of $T_c$.
 
To conclude, we should emphasize certain key points regarding importance of the mapping utilized here and its application to other spin glass models. First the mapping follows from a reorganization of the total phase space including the replica degrees of freedom without necessitating any ansatz of the $Q$ matrix structure (replica symmetric or otherwise) and thus only the ensemble of the disorder is chosen. Subsequently, within the CG representation, a direct confirmation of a thermodynamic phase transition can be observed without any knowledge of the proper order parameter (droplet or Parisi-like) of the original spin glass model. Furthermore, a virial expansion of the CG density allows in principle for the separation of mean-field (tree level diagrams) and correlation effects (loop diagrams) that may lead to replica symmetry breaking. Within the CG system of the short-ranged Ising spin glass systems in finite dimensions, the mean field theory should precisely recover the Sherrington-Kirkpatrick solution, and the correlation effects can be treated as perturbations which can affect the stability of the spin glass phase.

In the case of the SSGM, the facility in deriving the the CG model results from the integrability of the spin variables, however for Ising-like spins the methods presented here must be replaced by a different manipulation. Nevertheless, our emphasis is placed on the known rigorous mathematical understanding of replica based models, for which the trace over the total phase space can be generically re-expressed as some type of a CG ensemble properly taken in the $n\to0$ limit\cite{kanzieperPRL2002}. Thus, a successful application of this approach has been achieved here, and we look forward to developing the methods further in other spin glass models.

We are extremely grateful to Prof. P.J. Forrester of the University of Melbourne, Australia for sharing Ref.  \cite{FandW}, which is a crucial part of the analysis. Additionally, we give thanks to J. Landy for assistance and E. Kanzieper for references. S.A. is supported by a RIKEN FPR postdoctoral fellowship. J.R acknowledges support from the NSF through grant DMR 0704274.

\end{document}